\documentstyle[aps,prb,epsfig,floats]{revtex}
\begin{document}
\twocolumn[
\hsize\textwidth\columnwidth\hsize\csname@twocolumnfalse\endcsname
\draft
\title{Towards the fabrication of phosphorus qubits for a silicon quantum computer}
\author{J. L. O'Brien,$^{1,2}$\cite{email} S. R.
Schofield,$^{1,2}$ M. Y. Simmons,$^{1,2}$ R. G. Clark,$^{1,2}$ A.
S. Dzurak,$^{1,3}$ N. J. Curson,$^{1,2}$ B. E. Kane,$^4$ N. S.
McAlpine,$^2$\cite{nsm} M. E. Hawley,$^6$ G. W. Brown$^6$}
\address{$^1$Centre for Quantum Computer Technology, University of New South Wales, Sydney 2052, Australia}
\address{$^2$School of Physics, University of New South Wales, Sydney 2052, Australia}
\address{$^3$School of Electrical Engineering \& Telecommunications, University of New South Wales, Sydney 2052, Australia}
\address{$^4$Laboratory for Physical Sciences, University of Maryland, College Park, MD 20740, USA}
\address{$^5$Los Alamos National Laboratory, Los Alamos, NM 87545, USA}
\date{\today}
\maketitle
\begin{abstract}
The quest to build a quantum computer has been inspired by the
recognition of the formidable computational power such a device
could offer. In particular silicon-based proposals, using the
nuclear or electron spin of dopants as qubits, are attractive due
to the long spin relaxation times involved, their scalability,
and the ease of integration with existing silicon technology.
Fabrication of such devices however requires atomic scale
manipulation - an immense technological challenge. We demonstrate
that it is possible to fabricate an atomically-precise linear
array of single phosphorus bearing molecules on a silicon surface
with the required dimensions for the fabrication of a
silicon-based quantum computer. We also discuss strategies for the
encapsulation of these phosphorus atoms by subsequent silicon
crystal growth.\\({\em To appear in Phys. Rev. B Rapid Comm.})
\end{abstract}
]

A quantum bit (or qubit) is a two level quantum system that is
the building block of a quantum computer. To date the most
advanced realisations of a quantum computer are qubit ion trap
\cite{sa-nat-404-256} and nuclear magnetic resonance
\cite{ch-nat-393-143,ch-prl-80-3408,jo-prl-83-1050} systems.
However scaling these systems to large numbers of qubits will be
difficult \cite{wa-sci-277-1688}, making solid-state
architectures \cite{lo-pra-57-120}, with their promise of
scalability, important. In 1998 Kane proposed a novel solid state
quantum computer design \cite{ka-nat-393-133} using phosphorus
$^{31}$P nuclei (nuclear spin I = 1/2) as the qubits in
isotopically-pure silicon $^{28}$Si (I = 0). The device
architecture is shown in Fig. 1a, with phosphorus qubits embedded
in silicon approximately 20 nm apart. This separation allows the
donor electron wavefunctions to overlap, whilst an insulating
barrier isolates them from the surface control A and J gates.
These A and J gates control the hyperfine interaction between the
nuclear and electron spins and the coupling between adjacent
donor electrons respectively. For a detailed description of the
computer operation refer to Kane \cite{ka-nat-393-133}. An
alternative strategy using the electron spins of the phosphorus
donors as qubits has also been proposed \cite{vr-pra-62-012306}.

One of the major challenges of this design is to reliably
fabricate an atomically-precise array of phosphorus nuclei in
silicon - a feat that has yet to be achieved in a semiconductor
system. Whilst a scanning tunnelling microscope (STM) tip has
been used for atomic scale arrangement of metal atoms on metal
surfaces \cite{cr-sci-262-218}, rearrangement of individual atoms
in a semiconductor system is not straightforward due to the
strong covalent bonds involved. As a result, we have employed a
hydrogen resist strategy outlined in Fig. 1b. Here the array is
fabricated using a resist technology, much like in conventional
lithography, where the resist is a layer of hydrogen atoms that
terminate the silicon surface. An STM tip is used to selectively
desorb individual hydrogen atoms, exposing the underlying silicon
surface in the required array. STM induced hydrogen desorption
has been developed and refined over the past ten years
\cite{th-ss-411-203} and has been proposed \cite{tu-sse-42-1061}
for the assembly of atomically-ordered device structures. We
demonstrate a process to adsorb single phosphine molecules in a
predefined array, with atomic resolution, which we have developed
specifically for the fabrication of a silicon quantum computer.
The incorporation of these arrays in silicon is then discussed.

Whilst hydrogen lithography at the tens of Angstrom linewidth
scale has been used to selectively expose the silicon surface to
oxygen \cite{ly-jvstb-12-3735}, ammonia \cite{ly-jvstb-12-3735},
iron \cite{ad-jvstb-14-1642}, aluminium \cite{sh-prl-78-1271},
gallium \cite{ha-ss-386-161} and cobalt \cite{pa-jap-185-1907},
there has only been one recent report of adsorption, at the
atomic scale, in this case individual and clusters of silver
atoms \cite{sa-prb-62-16167}. Here we demonstrate the controlled
adsorption of a linear array of single phosphine molecules in the
extreme case of single hydrogen atom desorption sites for direct
application to the fabrication of a scalable silicon quantum
computer. This technical achievement has answered the critical
questions of whether a hydrogen resist is effective during
exposure to phosphine and whether or not phosphine will adsorb to
an STM-exposed site sufficiently small to achieve one and only
one PH$_3$ molecule at that site.

The requirements for this quantum computer design are very
stringent. In order to undertake high resolution lithography the
silicon surface must be atomically flat with a low defect density
to allow the formation of a near perfect resist layer, where one
hydrogen atom bonds to each surface silicon atom. The ability to
then desorb individual hydrogen atoms requires a sharp, large
cone angle tungsten tip in order to form $<$1 nm desorption
sites. These sites are subsequently exposed to high purity
phosphine gas for the required phosphorus atom placement (Fig.
1b). We demonstrate each of these steps below. Throughout this
process particular attention must be made to avoid the
introduction of any spin or charge impurities that would be fatal
to the operation of the quantum computer \cite{ka-nat-393-133}.

\begin{figure}
\begin{center}
\vspace{-.4cm}
\includegraphics*[width=7cm]{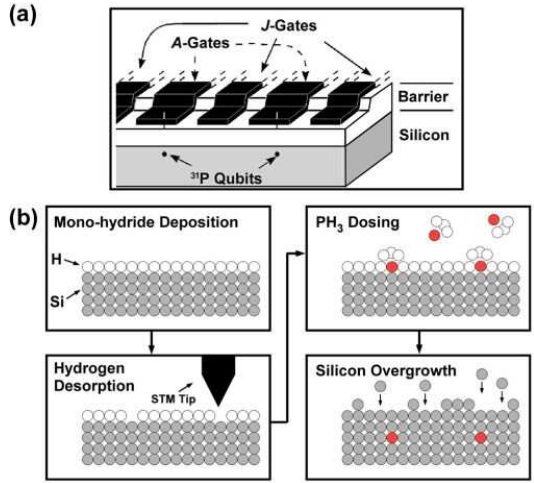}
\end{center}
\caption{A schematic of the process to fabricate the Kane
architecture. (a) Detail of the Kane quantum computer
architecture taken from Ref. \protect\onlinecite{ka-nat-393-133},
showing two phosphorus qubits in a linear array, incorporated into
isotopically-pure $^{28}$Si and isolated from surface metal A and
J gates by an insulating barrier. (b) A process to fabricate an
array of phosphorus qubits in silicon. A low defect density
Si(001)2x1 surface is passivated with a monolayer of hydrogen. An
STM tip is used to selectively desorb hydrogen, exposing silicon
on an atomic scale permitting only one phosphine molecule to
adsorb at each of the required sites. Low temperature silicon
overgrowth encapsulates the phosphorus.} \vspace{-.3cm}
\label{fig1}
\end{figure}

The final fabrication step outlined in Fig. 1b which is not the
subject of this paper, but presents a significant challenge, is
to encapsulate the phosphorus qubits in a crystalline lattice of
isotopically-pure silicon-28 \cite{br-prl-81-393}. The main
difficultly in this step is to ensure the phosphorus atoms
incorporate into the silicon crystal and remain in their ordered
atomic array. The most direct route to achieving this is to desorb
the hydrogen resist from the surface by heating to $\sim$700 K
followed by epitaxial silicon growth over the phosphorus array. A
possible concern with heating the surface is the potential to
induce lateral surface diffusion of the phosphorus atoms in the
array. However it is known that the phosphorus atom in the PH$_2$
molecule, with a single bond to the silicon, incorporates into
the silicon surface with a more stable threefold coordination
geometry over the lower temperature range of 550-650 K
\cite{li-prb-61-2799}. We can avoid heating the surface at all
during hydrogen desorption by either direct optical excitation at
$\lambda$ = 157 nm \cite{vo-prl-82-1967} or electron bombardment
\cite{gu-ss-62-239}. An alternative room temperature approach
that will help to incorporate the phosphorus into the silicon
lattice without significant diffusion involves photo-induced
excitation of the attached PH$_2$ molecule to increase its
chemical activity with the surface \cite{gu-ss-62-239}. Finally
it is also possible to leave the hydrogen resist and grow
epitaxial silicon directly on the mono-hydride surface
\cite{co-prl-72-1236}. In all above cases 10-30 {\AA} of epitaxial
silicon can be grown at low temperatures, down to room
temperature \cite{ea-prl-65-1227} to encapsulate the phosphorus,
followed by elevated temperature growth at $\sim$500 K to maintain
crystallinity \cite{ea-prl-65-1227}. We will return to the issue
of phosphorus incorporation in our final discussion.

\begin{figure}
\vspace{-.4cm}
\begin{center}
\includegraphics*[width=6cm]{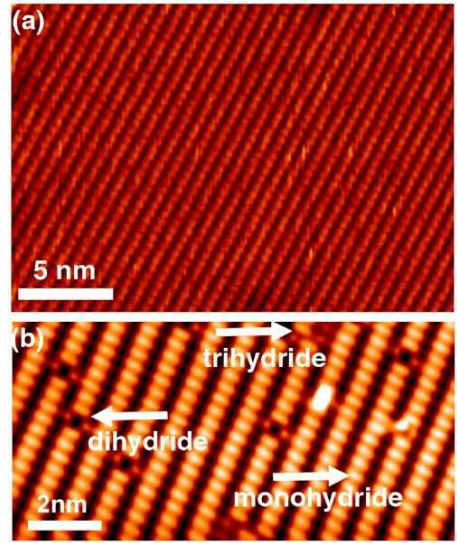}
\end{center}
\caption{Preparation of clean and hydrogen terminated surfaces.
(a) Filled state STM image of a clean, very low defect density
Si(001)2x1 surface. This image was acquired at a sample bias of
-1.0 V and a tunnelling current of 0.4 nA. (b) Fully hydrogen
terminated Si(001)2x1 surface. Image acquired at -2.5 V and 0.4
nA. The surface is almost entirely mono-hydride, with sites of
di-hydride and tri-hydride also identified.} \vspace{-.3cm}
\label{fig2}
\end{figure}

The first step in fabricating a phosphorus qubit array is to
prepare a clean, defect free silicon surface. Fig. 2a demonstrates
an optimally-prepared Si(001)2$\times$1 surface consisting of rows
of $\sigma$-bonded silicon-silicon dimers. By heating the surface
to 1200 $^{\circ}$C in a variable-temperature ultra-high vacuum
STM system and performing a controlled cool-down
\cite{sw-jvsta-7-2901} we have achieved large defect free
regions. The ``bean'' shaped protrusions in this filled state
image correspond to the charge overlap of the electrons in the
dangling bonds on each silicon surface atom giving rise to a weak
$\pi$-bond. These dangling bonds make the surface reactive,
allowing the subsequent adsorption of species such as hydrogen
and phosphine.

\begin{figure}
\vspace{-.4cm}
\begin{center}
\includegraphics*[width=7cm]{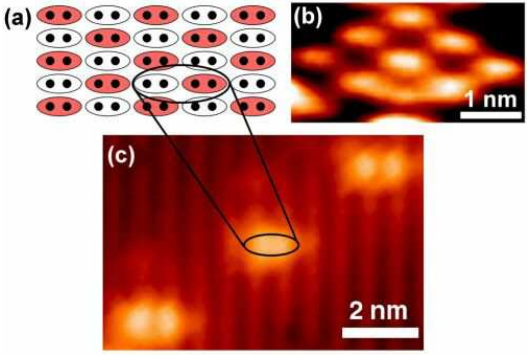}
\end{center}
\caption{Bonding structure of phosphine on Si(001)2x1. (a)
Schematic of the c(4$\times$2) structure of PH$_3$ bonded to the
Si(001)2x1 surface, where red dimers indicate PH$_3$ bonding
sites. (b) An STM image of this structure, acquired at a sample
bias of -3.0 V and tunnelling current 0.2 nA. (c) STM image of
three hydrogen desorption sites on a mono-hydride Si(001)2x1:H
surface. The highlighted regions in (a) and (c) indicate that only
one phosphine molecule can adsorb at each desorption site.}
\vspace{-.3cm} \label{fig3}
\end{figure}

The adsorption of atomic hydrogen on the Si(001)2$\times$1
surface can occur in three ways depending on the surface
temperature, forming either a mono-, di- or tri-hydride
\cite{ou-ssr-35-1}. An atomic hydrogen source consisting of a
heated tungsten filament and water-cooled heat shroud has been
used for this work. Fig. 2b shows an image taken at room
temperature of a low defect density Si(001)2$\times$1 surface
after exposure to high purity atomic hydrogen at 600 K. It can
clearly be seen that a near uniform coverage of the mono-hydride
phase has occurred, where one hydrogen atom bonds to each silicon
atom. Sites of di-hydride, where two hydrogen atoms bond to each
silicon, and tri-hydride, a mixture of the other two phases, are
also indicated. We have found that all three phases passivate the
surface and act as an effective resist during subsequent exposure
to phosphine. Comparison of current-voltage spectroscopy before
and after hydrogen dosing (not shown) confirms the existence of a
hydrogen passivated surface, with the silicon $\pi$*-antibonding
peak evolving into the silicon-hydrogen antibonding peak after
hydrogenation \cite{ha-prl-59-2071}.

\begin{figure}[t]
\vspace{-0.4cm}
\begin{center}
\includegraphics*[width=6cm]{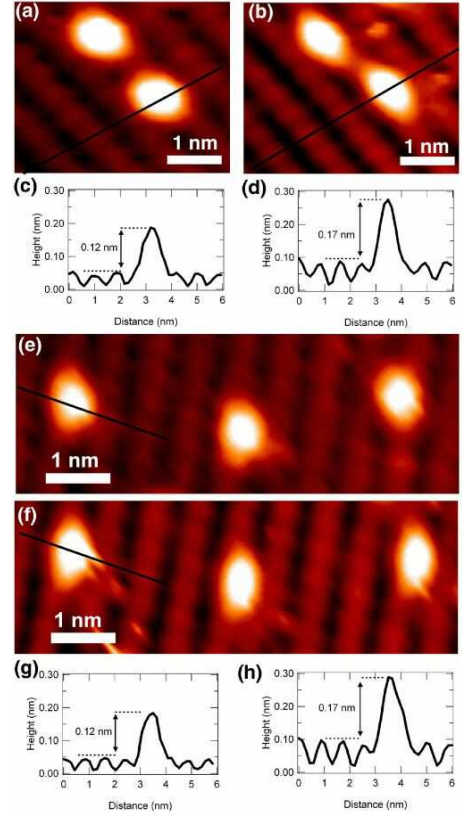}
\end{center}
\caption{Demonstration of single PH$_3$ adsorption through a STM
patterned hydrogen resist. STM images of two desorption sites
before (a) and after (b) phosphine dosing. The corresponding line
profiles (c and d) show a characteristic height increase of
$\sim$0.05 nm. Three desorption sites before (e) and after (f)
phosphine dosing and corresponding line profiles (g and h). All
images were acquired at a sample bias of -1.8 V and tunnelling
current of 0.4 nA.} \vspace{-.3cm} \label{fig4}
\end{figure}

An STM tip is then used to desorb hydrogen from the surface by
application of a controlled voltage pulse between the tip and
sample \cite{th-ss-411-203}. Optimisation of the geometry of an
oxide-free tungsten tip and controlled voltage pulses (sample
bias $\sim$6 V and tunnelling current $\sim$1 nA for $\sim$1 ms)
makes atomic resolution desorption possible. In order to allow
the adsorption of one phosphine molecule, and therefore only one
phosphorus atom, it is necessary to desorb an area that exposes
less than or equal to two silicon dimers as shown schematically
in Fig. 3a. This is because phosphine bonds to the
Si(001)2$\times$1 surface with a c(4$\times$2) surface
periodicity as demonstrated in Fig. 3b where we have dosed a clean
Si(001)2$\times$1 surface with phosphine \cite{li-prb-61-2799}.
The STM image in Fig. 3c shows three $<$ 1 nm diameter hydrogen
desorption sites in a row with a pitch of $\sim$4 nm on a hydrogen
terminated Si(001)2$\times$1 surface. This image, with such a
close spacing between sites, highlights the atomic resolution
desorption achieved. The distance between sites can easily be
increased to the required qubit spacing of 20 nm, and we have
performed controlled lithography of single desorption sites in a
line $>$ 100 nm in length. The desorption sites in Fig. 3c appear
as bright protrusions as a result of the extension of electron
density out of the surface due to the silicon dimer surface
states of the exposed silicon dangling bonds
\cite{th-ss-411-203}. Fig. 3 demonstrates that these desorption
sites are sufficiently small to allow only one phosphine molecule
to bond to the surface at each site.

\begin{figure}
\vspace{-.4cm}
\begin{center}
\includegraphics*[width=6cm]{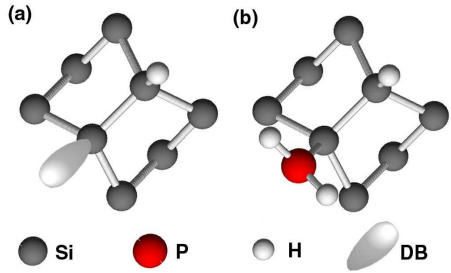}
\end{center}
\caption{Phosphine adsorption \protect\nolinebreak on the
patterned Si(001)2x1:H surface. Schematic of a single silicon
dangling bond (DB) (a) and the Si-PH$_2$ structure expected after
dissociation of the adsorbed PH$_3$ (b). These models are shown
looking down onto the surface, where the two silicon atoms in the
centre are surface atoms and the others are sub-surface atoms.}
\vspace{-.3cm} \label{fig5}
\end{figure}

Following STM lithography to expose small regions of the
Si(001)2$\times$1 surface it is then necessary to adsorb the
individual phosphorus atoms or qubits using phosphine gas. Figs.
4a and 4b show atomic resolution images of two hydrogen
desorption sites $\sim$1.5 nm apart on a dimer row both before
and after exposure to phosphine gas. The effectiveness of the
hydrogen resist as a barrier to phosphine adsorption is
demonstrated by the uniform hydrogen coverage after phosphine
dosing except at the previously desorbed hydrogen sites. In order
to observe any changes after phosphine exposure we have
specifically created single hydrogen desorption sites, rather
than larger desorption sites (as in Fig. 3) and present high
resolution images where the spacing between sites is very small.
Here the bright protrusion at each of the desorption sites in
Fig. 4a is the signature of the single silicon dangling bond,
after desorption of just one hydrogen atom, in this case from the
left side of the dimer. The remaining hydrogen on the silicon
dimer is known to be transient \cite{hi-prb-60-15896} and we have
observed it diffusing from one side of the dimer to the other.

In order to obtain high purity phosphine gas delivery, the PH$_3$
micro-dosing system and its connection to the UHV STM employed
internally electro-polished gas lines assembled in a clean-room
environment. Mass spectra taken in the chamber during the
exposure at a pressure of 10$^{-8}$ mbar reveal no significant
increase in the partial pressure of any other species. The
sticking coefficient of phosphine on the clean silicon surface is
1 \cite{li-prb-61-2799}. Fig. 4b shows the same area as Fig. 4a
after exposure to phosphine gas at room temperature. Analysis of
the line profiles in Fig. 4c and 4d shows a characteristic
increase of $\sim$0.05 nm in the height of the protrusion after
phosphine dosing \cite{note}. Figs. 4d and 4e show three
desorption sites, in a line perpendicular to the dimer rows,
before and after phosphine dosing. The associated line profiles
(Figs. 4f and 4g) again show a $\sim$0.05 nm height increase.
This increase, observed at all adsorption sites over several
images, was calibrated against an atomic step edge on the same
surface (not shown) both before and after phosphine dosing. This
reproducible increase confirms the adsorption of a PH$_3$ molecule
and corresponds to the difference between the exposed silicon
dangling bond and the adsorbed phosphine. The transient nature of
the hydrogen atom on the silicon dimer can account for the
asymmetry of the image in Fig. 4b where one phosphine molecule
has bonded to the left silicon in the dimer (upper) and another
phosphine molecule has bonded to the right silicon in the dimer
(lower).

Studies of the interaction of phosphine with the clean
Si(001)2$\times$1 surface \cite{li-prb-61-2799} suggest that
PH$_3$ molecularly adsorbs to one end of a silicon dimer and can
then dissociate to PH$_2$ provided enough silicon dangling bonds
are available nearby for the re-adsorption of the dissociated
hydrogen. The absence of available dangling bond sites on the
hydrogen-terminated surface inhibits this dissociation step. Fig.
5a shows a schematic of a single dangling bond site on a hydrogen
terminated surface before phosphine dosing and Fig. 5b shows the
proposed structure of Si-PH$_2$ after dissociation of the
adsorbed PH$_3$. In this configuration the adsorbed phosphorus
atom maintains the tetrahedral geometry and threefold
coordination, with the PH$_2$ bonded to one end of the dimer
\cite{li-prb-61-2799}. A similar dissociative process and
Si-XH$_2$ configuration is understood to occur in the adsorption
of NH$_3$ \cite{zh-jcp-112-244} and AsH$_3$ \cite{no-prb-51-2218}
on the Si(001)2$\times$1 surface, where both nitrogen and arsenic
are isoelectronic with phosphorus.

In the above discussion we have demonstrated the process of
controlled single phosphine molecule adsorption for the extreme
case of single hydrogen atom desorption sites. In future work it
will be important to maintain the ordered atomic array during the
final silicon growth step (final panel of Fig. 1b). Here the
optimisation of the size and crystallographic orientation of the
desorption sites will be critical. Dissociation of singly-bonded
PH$_2$ to threefold coordinate P + 2H can proceed above 500 K
\cite{ea-prl-65-1227}, provided there are suitable dangling bond
sites that overcome competition with the recombinative desorption
of PH$_3$.  An alternative option is to dose the
hydrogen-patterned surface with phosphine gas at elevated
temperatures ($\sim$550 K) to directly induce phosphorus
incorporation into the first atomic layer. We have independently
performed elevated temperature studies which indicate that at
these temperatures, the silicon surface is stable.

We have demonstrated the effectiveness of a hydrogen resist as a
barrier to phosphine adsorption and used STM lithography for the
controlled placement of single phosphorus bearing molecules on a
silicon surface - central to the construction of a silicon based
quantum computer. This process, shown for closely spaced
controlled doping, demonstrates the achievability of more
widely-spaced ($\sim$20 nm), precisely positioned phosphorus
qubit arrays over large areas. Whilst it is not yet possible to
guarantee the subsequent overgrowth steps required for the Kane
architecture, this letter demonstrates a significant advance in
the solid-state qubit fabrication process and bodes well for the
realisation of a scalable silicon quantum computer. Finally, it
is important to note that the fabrication strategy demonstrated
here is also directly applicable to other silicon based quantum
computer architectures \cite{vr-pra-62-012306}.

We would like to thank I. Andrienko, T. Kamins, G. J. Milburn, S.
Prawer, C. Thirstrup and  S. Williams for useful discussions.
This work is funded by the Australian Research Council Special
Research Centre scheme, the Australian Government, the US National
Security Agency, and the US Advanced Research and Development
Activity.


\begin{thebibliography}{10}

\bibitem[*]{email}e-mail: job@phys.unsw.edu.au
\bibitem[{\dag}]{nsm}present address: School of Mathematical and Physical Sciences, University of Newcastle, Callaghan, New South Wales, 2308, Australia

\bibitem{sa-nat-404-256} C. A. Sackett, et al., Nature
\textbf{404}, 256 (2000).

\bibitem{ch-nat-393-143} I. L. Chuang, L. M. K. Vandersypen,
Xinlan Zhou, D. W. Leung, S. Lloyd, Nature \textbf{393}, 143
(1998).

\bibitem{ch-prl-80-3408} I. L. Chuang, N. Gershenfeld, M.
Kubinec, Phys. Rev. Lett. \textbf{80}, 3408 (1998).

\bibitem{jo-prl-83-1050} J. A. Jones, M. Mosca, Phys. Rev. Lett.
\textbf{83}, 1050 (1999).

\bibitem{wa-sci-277-1688} W. S. Warren, Science \textbf{277},
1688 (1997).

\bibitem{lo-pra-57-120} D. Loss, D. P. DiVincenzo, Phys. Rev. A
\textbf{57}, 120 (1998).

\bibitem{ka-nat-393-133} B. E. Kane, Nature \textbf{393}, 133
(1998).

\bibitem{vr-pra-62-012306} R. Vrijen, et al., Phys. Rev. A
\textbf{62}, 012306 (2000).

\bibitem{cr-sci-262-218} M. F. Crommie, C. P. Lutz, D. M.
Eigler, Science \textbf{262}, 218 (1993).

\bibitem{th-ss-411-203} C. Thirstrup, M. Sakurai, T. Nakayama,
M. Aono, Surf. Sci., \textbf{411}, 203 (1998) and references
therein.

\bibitem{tu-sse-42-1061} J. R. Tucker, T. -C. Shen, Solid State
Electronics \textbf{42}, 1061 (1998).

\bibitem{ly-jvstb-12-3735} J. W. Lyding, G. C. Abeln, T. -C.
Shen, C. Wang, J. R. Tucker, J. Vac. Sci. Technol. B \textbf{12},
3735 (1994).

\bibitem{ad-jvstb-14-1642} D. P. Adams, T. M. Mayer, B. S.
Swartzentruber, J. Vac. Sci. Technol. B \textbf{14}, 1642 (1996).

\bibitem{sh-prl-78-1271} T. -C. Shen, C. Wang, J. R. Tucker,
Phys. Rev. Lett. \textbf{78}, 1271 (1997).

\bibitem{ha-ss-386-161} T. Hashizume, S. Heike, M. I. Lutwyche,
S. Watanabe, Y. Wada, Surf. Sci. \textbf{386}, 161 (1997).

\bibitem{pa-jap-185-1907} G. Palasantzas, B. Ilge, J. De Nijs,
L. J. Geerligs, J. Appl. Phys. \textbf{185}, 1907 (1999).

\bibitem{sa-prb-62-16167} M. Sakurai, C. Thirstrup, M. Aono,
Phys. Rev. B. \textbf{62}, 16167 (2000).

\bibitem{br-prl-81-393} H. Bracht, E. E. Haller, R.
Clark-Phelps, Phys. Rev. Lett. \textbf{81}, 393 (1998)

\bibitem{li-prb-61-2799} D. S. Lin, T. S. Ku, R. P. Chen, Phys.
Rev. B \textbf{61}, 2799 (2000) and references therein.

\bibitem{vo-prl-82-1967} T. Vondrak, X. -Y. Zhu, Phys. Rev.
Lett. \textbf{82}, 1967 (1999).

\bibitem{gu-ss-62-239} H. Guo, P. Saalfrank, T. Seideman, Prog.
Surf. Sci. \textbf{62}, 239 (1999)

\bibitem{co-prl-72-1236} M. Copel, R. M. Tromp, Phys. Rev. Lett.
\textbf{72}, 1236 (1994)

\bibitem{ea-prl-65-1227} D. J. Eaglesham, H. -J. Gossmann, M.
Cerullo, Phys. Rev. Lett.  \textbf{65}, 1227 (1990).

\bibitem{sw-jvsta-7-2901} B. S. Swartzentruber, Y. W. Mo, M. B.
Webb, M. G. Lagally, J. Vac. Sci. Tech. A \textbf{7}, 2901 (1989).

\bibitem{ou-ssr-35-1} K. Oura, V. G. Lifshits, A. A. Saranin, A.
V. Zotov, M. Katayama, Surf. Sci. Rep. \textbf{35}, 1 (1999).

\bibitem{ha-prl-59-2071} R. J. Hamers, Ph. Avouris, F. Bozso,
Phys. Rev. Lett. \textbf{59}, 2071 (1987).

\bibitem{hi-prb-60-15896} E. Hill, B. Freelon, E. Ganz, Phys.
Rev. B \textbf{60}, 15896 (1999).

\bibitem{note} A slightly larger height in the line profile of
the dimer rows is observed after PH$_3$ dosing. Such a difference
can frequently occur due to minor changes in imaging conditions
between scans, which results in the STM tip extending further
into the gap between dimer rows. However the height difference
due to PH$_3$ adsorption is measured from the top of the dimer
rows to the top of the protrusion and is not therefore affected by
this.

\bibitem{zh-jcp-112-244} Zhi-Heng Loh, H. C. Kang, J. Chem.
Phys. \textbf{112}, 2444 (2000) and references therin.

\bibitem{no-prb-51-2218} J. E. Northrup Phys. Rev. B
\textbf{51}, 2218 (1995).

\end{thebibliography}
\end{document}